\documentclass[prd,twocolumn,showpacs,preprintnumbers,amsmath,amssymb]{revtex4}

\usepackage{amsmath}
\usepackage{graphicx}
\usepackage{dcolumn}
\usepackage{xspace}
\usepackage{color}
\usepackage{url}
\usepackage[utf8]{inputenc}
\usepackage{epstopdf}

\allowdisplaybreaks

\newcommand{\bea}{\begin{eqnarray}}
\newcommand{\eea}{\end{eqnarray}}

\newcommand{\eq}[1]{Eq.~\eqref{#1}}

\begin{document}


\preprint{\hbox{PSI-PR-18-15}}

\title{\boldmath QCD Improved Matching for Semi-Leptonic $B$ Decays with Leptoquarks}

\author{Jason Aebischer}
\email{jason.aebischer@tum.de}
\affiliation{Excellence Cluster Universe, Boltzmannstr. 2, 85748 Garching, Germany}

\author{Andreas Crivellin}
\email{andreas.crivellin@cern.ch}
\affiliation{Paul Scherrer Institut, CH--5232 Villigen PSI, Switzerland}

\author{Christoph Greub}
\email{greub@itp.unibe.ch}
\affiliation{Albert Einstein Center for Fundamental Physics, Institute for Theoretical Physics, University of Bern, CH-3012 Bern,Switzerland}



\begin{abstract}

Leptoquarks (LQs) provide very promising solutions to the tensions between the experimental measurements and the SM predictions of $b\to s\ell^+\ell^-$ and $b\to c\tau\nu$ processes. In this case the LQ masses are in general at the TeV scale and they can thus be produced at high energy colliders and dedicated LHC searches are ongoing. While for LQ production and decay the $O(\alpha_s)$ corrections have been known for a long time, the $O(\alpha_s)$ corrections to the matching on 2-quark-2-lepton operators have not been calculated, yet. In this article we close this gap by computing the QCD corrections to the matching of LQ models on the effective SM Lagrangian for both scalar and vector LQs. We find an enhancement of the Wilson coefficients of vector operators with respect to the tree-level results of around 8 \% (13 \%) if they originate from scalar (vector) LQs. This softens the LHC bounds and increases the allowed parameter space of LQ models addressing the flavour anomalies.

\end{abstract}




\maketitle


\section{Introduction}

\label{intro}

Significant deviations from the SM predictions in $b\to s\mu^+\mu^-$ processes (above the $5\,\sigma$ level \cite{Capdevila:2017bsm}\footnote{Including only $R(K)$ and $R(K^*)$ the significance is at the $4\,\sigma$ level~\cite{Altmannshofer:2017yso,DAmico:2017mtc,Geng:2017svp,Ciuchini:2017mik,Hiller:2017bzc,Hurth:2017hxg}.}) and in $b\to c\tau\nu$  processes (at the 4~$\sigma$ level \cite{Amhis:2016xyh}) were observed in recent years. These observations strongly point towards the violation of lepton flavour universality in semileptonic $B$ decays, suggesting a possible connection between these two classes of decays. In this context leptoquarks\footnote{See Ref.~\cite{Dorsner:2016wpm} for a recent review.} (LQs) are natural candidates for an explanation, since they give tree-level effects to semi-leptonic processes while their contributions to other flavour observables (which in general agree very well with the SM) are loop-suppressed. In fact, LQs (including squarks in the R-parity violating MSSM) have been extensively employed to explain the anomalies in $b\to s\mu^+\mu^-$~\cite{Gripaios:2014tna,Biswas:2014gga,Sahoo:2015wya,Huang:2015vpt,Varzielas:2015iva,Pas:2015hca,Deppisch:2016qqd,Chen:2016dip,Becirevic:2016oho,Hiller:2016kry,Duraisamy:2016gsd,Barbieri:2016las,Becirevic:2017jtw,Guo:2017gxp,Aloni:2017ixa,Cline:2017aed,Fajfer:2018bfj,Sahoo:2018ffv,Hati:2018fzc,deMedeirosVarzielas:2018bcy} or $b\to c\tau\nu$ \cite{Deshpande:2012rr,Tanaka:2012nw,Sakaki:2013bfa,Freytsis:2015qca,Hati:2015awg,Li:2016vvp,Zhu:2016xdg,Popov:2016fzr,Deshpand:2016cpw,Altmannshofer:2017poe,Kamali:2018fhr,Azatov:2018knx,Wei:2018vmk,Hu:2018lmk} processes. Furthermore, they can even provide a common explanation~\cite{Bhattacharya:2014wla,Alonso:2015sja,Calibbi:2015kma,Fajfer:2015ycq,Greljo:2015mma,Barbieri:2015yvd,Bauer:2015knc,Boucenna:2016qad,Das:2016vkr,Bhattacharya:2016mcc,Becirevic:2016yqi,Sahoo:2016pet,Crivellin:2017zlb,Cai:2017wry,Chen:2017hir,Dorsner:2017ufx,Buttazzo:2017ixm,DiLuzio:2017vat,Bordone:2017bld,Barbieri:2017tuq,Calibbi:2017qbu,Blanke:2018sro,Greljo:2018tuh,Bordone:2018nbg,Crivellin:2018yvo,Marzocca:2018wcf,Aydemir:2018cbb,Fayyazuddin:2018zww,Matsuzaki:2018jui,Alvarez:2018gxs,Becirevic:2018afm,Kumar:2018kmr,Azatov:2018kzb,DiLuzio:2018zxy,Faber:2018qon,Biswas:2018snp,Angelescu:2018tyl,Choi:2018stw,Heeck:2018ntp}\footnote{Leptoquarks have also been discussed in the context of $\varepsilon'/\varepsilon$ and rare Kaon decays in Ref.~\cite{Bobeth:2017ecx} and for electric dipole moments in Ref.~\cite{Dekens:2018bci}.}.

Direct searches for LQs at the LHC have been performed~\cite{Aaboud:2017faq,Aaboud:2017nmi,Sirunyan:2017yrk,Sirunyan:2017dhe,CMS-PAS-EXO-17-029} and also projections for future colliders in the context of the above mentioned flavour anomalies have been investigated~\cite{Allanach:2017bta}. For collider processes the QCD corrections to production and decay of LQs are known for a long time~\cite{Plehn:1997az,Kramer:1997hh,Kramer:2004df} and have been improved to include NLO parton shower~\cite{Mandal:2015lca} or a large width~\cite{Hammett:2015sea}. Furthermore, in recent analyses correlating the $B$ anomalies to LHC searches~\cite{Faroughy:2016osc,Dorsner:2017ufx,Dorsner:2018ynv,Hiller:2018wbv,Monteux:2018ufc,Schmaltz:2018nls} QCD corrections to production and/or decay were included. However, the analogous $\alpha_s$ corrections for LQ effects in the low energy observables (i.e. semi-leptonic $B$ decays), which should be taken into account for consistency, are still missing.

In this article we therefore compute the 1-loop QCD corrections to the matching of models with LQs on the effective 4-fermion SM Lagrangian. After establishing our conventions in the next section, we perform the computation both for scalar and vector leptoquarks in a general gauge for the gluon fields in Sec.~\ref{secCalc}. Finally we examine the importance of the calculated effects and conclude.

\begin{figure*}[t]
	\begin{center}
		\begin{tabular}{cp{7mm}c}
			\includegraphics[width=0.9\textwidth]{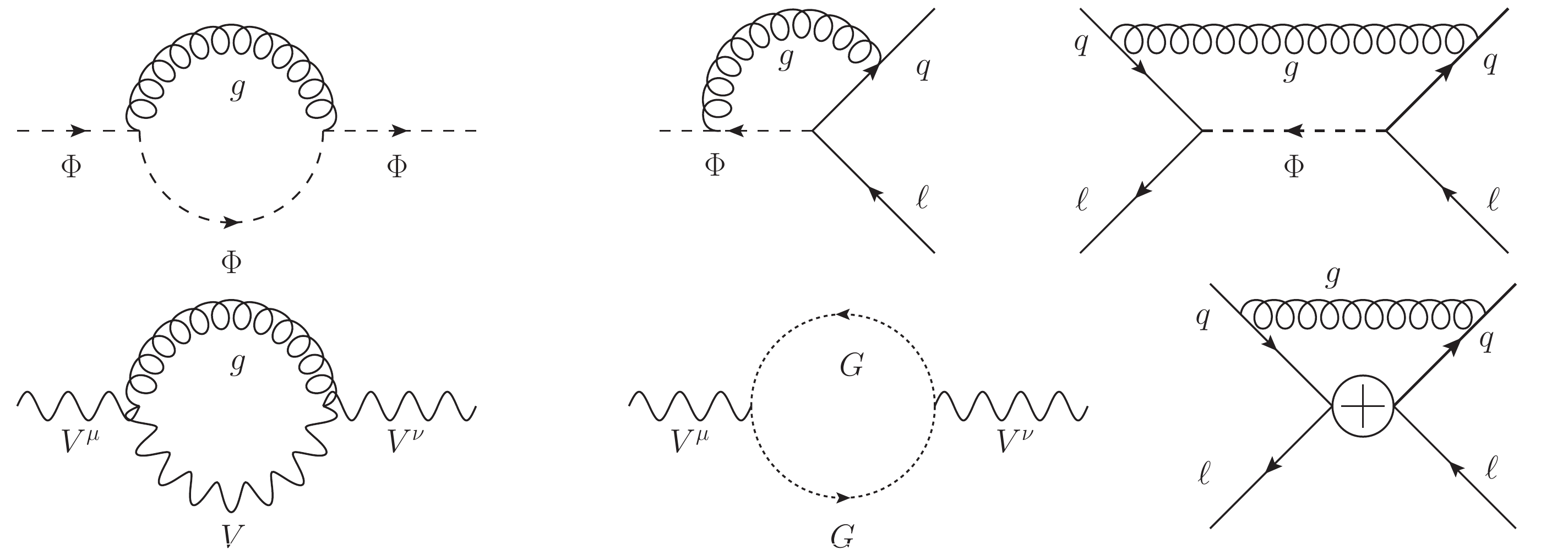}
		\end{tabular}
	\end{center}
	\caption{Examples of Feynman diagrams in the full and in the effective  theory. $G$ denotes the ghost contribution. Not shown are the vector LQ self-energy with a Goldstone and a gluon as well as the vertex correction for the VLQ. }
	\label{Feynman}
\end{figure*}

\section{Setup}

As a starting point we consider the following generic Lagrangian governing the couplings of scalar (vector) LQs $\Phi$ ($V_\mu$) of mass $M$ to leptons $\ell$ (charged leptons or neutrinos) and quarks $q$ (up or down type):

\begin{eqnarray}
\begin{aligned}
L_{q\ell}^{LQ} =& \bar q\left( {{\Gamma_L^S}{P_L} + {\Gamma_R^S}{P_R}} \right)\ell {\Phi ^*} + {\rm{h}}{\rm{.c}}{\rm{.}}\\
& + \bar q\left( {{\Gamma_L^V}\gamma^\mu{P_L} + {\Gamma_R^V}\gamma^\mu{P_R}} \right)\ell {V^*_\mu} + {\rm{h}}{\rm{.c}}{\rm{.}}\,.
\label{LLQ}
\end{aligned}
\end{eqnarray}
Note that here we do not consider gauge invariance with respect to $SU(2)_L$ or $U(1)_Y$. This is possible for our purpose since these gauge symmetries are disjunct from $SU(3)_c$. We also do not explicitly include the possibility of charge conjugated fields because this again does not affect the calculation of the QCD corrections. We will come back to the issue of charge conjugation later when we discuss the phenomenological importance of our results.

Let us now define the effective Lagrangian containing only SM fields in the $\bar q\ell\bar\ell q$ basis, which we call ``LQ basis''
\begin{eqnarray}
\begin{aligned}
L_{{\rm{eff}}}^{{\rm{LQ}}} &= \widetilde C_S^{AB}\widetilde O_S^{AB} + \widetilde C_V^{AB}\widetilde O_V^{AB} +\widetilde C_T^{A}\widetilde O_T^{A}\,,\\
\widetilde O_S^{AB} &= \bar q{P_A}\ell \bar \ell {P_B}q\,,\\
\widetilde O_V^{AB} &= \bar q{\gamma ^\mu }{P_A}\ell \bar \ell {\gamma _\mu }{P_B}q\,,\\
\widetilde O_T^A &= \bar q{\sigma ^{\mu \nu }}{P_A}\ell\bar \ell {\sigma _{\mu \nu }}{P_A}q\,,
\label{LQeffL}
\end{aligned}
\end{eqnarray}
as well as the corresponding operators in the ``SM basis'' with $\bar q q\bar\ell\ell$ operators
\begin{eqnarray}
\begin{aligned}
L_{{\rm{eff}}}^{{\rm{SM}}} &= C_S^{AB}O_S^{AB} + C_V^{AB}O_V^{AB} + C_T^AO_T^A\,,\\
O_S^{AB} &= \bar q{P_A}q\bar \ell {P_B}\ell\, ,\\
O_V^{AB} &= \bar q{\gamma ^\mu }{P_A}q\bar \ell {\gamma _\mu }{P_B}\ell ,\;\;\\O_T^A &= \bar q{\sigma ^{\mu \nu }}{P_A}q\bar \ell {\sigma _{\mu \nu }}{P_A}\ell\,.
\label{LSM}
\end{aligned}
\end{eqnarray}

Here $A,B=L,R$ label the chiralities. Performing the tree-level matching we obtain in the LQ basis
\begin{eqnarray}
\begin{aligned}
\widetilde C_S^{LR} &= \frac{{{{\left| {{\Gamma_L^S}} \right|}^2}}}{{{M^2}}},\\
\widetilde C_S^{LL} &= \frac{{\Gamma _L^S}\Gamma_R^{S*}}{{{M^2}}}\,,\\
\widetilde C_T^L&=0\,,\\
\widetilde C_V^{LL} &= -\frac{{{{\left| {{\Gamma_L^V}} \right|}^2}}}{{{M^2}}}\,,\\
\widetilde C_V^{LR} &= -\frac{{{\Gamma_L^V}\Gamma_R^{V*}}}{{{M^2}}}\,,
\label{WCtree}
\end{aligned}
\end{eqnarray}
and the corresponding formula with $L \leftrightarrow R$. Using standard Fierz identities (see e.g.~\cite{Fierz1937,Nieves:2003in})
\begin{eqnarray}
\begin{aligned}
C_V^{LL} &= \widetilde C_V^{LL} \!\!\! &=& -\frac{{{{\left| {{\Gamma_L^V}} \right|}^2}}}{{{M^2}}}\,,\\
C_S^{LR} &=  - 2\widetilde C_V^{RL}  \!\!\!&=& 2\frac{{\Gamma _{R}^V\Gamma _{L}^{V*}}}{{{M^2}}}\,,\\
C_T^{L} &=  - \frac{1}{8}\widetilde C_S^{LL} \!\!\! &=& -   \frac{{  1}}{8}\frac{{\Gamma _{L}^S\Gamma _{R}^{S*}}}{{{M^2}}}\,,\\
C_S^{LL} &=  - \frac{1}{2}\widetilde C_S^{LL} \!\!\! &=&  - \frac{1}{2}\frac{{\Gamma _{L}^S\Gamma _{R}^{S*}}}{{{M^2}}}\,,\\
C_V^{LR} &=  - \frac{1}{2}\widetilde C_S^{RL}  \!\!\!&=&  - \frac{1}{2}\frac{{{{\left| {\Gamma _{R}^S} \right|}^2}}}{{{M^2}}}\,,
\end{aligned}
\label{tree}
\end{eqnarray}
where we again do not show the results which are obtained by an interchange of chiralities $L \leftrightarrow R$.

\section{Calculation and Results}
\label{secCalc}

Let us now turn to the calculation of the QCD corrections to the
Wilson coefficients. Here, the same procedure is applied as within the SM when integrating out the $W$ boson~\cite{Buchalla:1995vs} in order to determine the
$\alpha_s$ corrections to \eq{tree}. We performed the calculation, also with the help of FeynArts~\cite{Hahn:2000kx} and FeynCalc~\cite{Mertig:1990an}, in dimensional regularization with naive anti-commuting $\gamma_5$.

We assume that the vector LQ (VLQ) is a gauge boson of an unspecified gauge group. Thus, its couplings (and the ones of the corresponding Goldstone  bosons) to gluons and ghosts are determined uniquely by requiring $SU(3)_c$  gauge invariance and the corresponding Lagrangian which also contains the  mass term of the VLQ with mass $M$ is given by~\cite{Blumlein:1996qp}
\begin{equation}
{L_{\rm VLQ}^{\rm QCD}} =
- \frac{1}{2}K_{\mu \nu }^{\alpha \dag }K_\alpha ^{\mu \nu } +
i{g_s}V_\mu ^{\alpha \dag }T_{\alpha \beta }^aV_\nu ^{\beta }G_a^{\mu \nu
} + M_{}^2 V_\mu ^{\alpha \dag }V_\alpha ^{\mu }\,,
\end{equation}
with \begin{equation}
K_{\mu \nu }^\alpha  = \left( {D_\mu ^{\alpha \beta }V_{\nu \beta }^{} -    D_\nu ^{\alpha \beta }V_{\mu \beta }^{}} \right)\,.\end{equation}
Here, $D_\mu ^{\alpha
  \beta } = {\partial _\mu }{\delta ^{\alpha \beta }} + i{g_s}T_a^{\alpha
  \beta }A_\mu ^a$
 is the covariant derivative with respect to QCD, $\alpha$ and $\beta$ are colour indices and $a$ labels the eight generators $T_a$ of $SU(3)_c$, $A_\mu ^a$ are the gluon fields and $G_a^{\mu\nu}$ is the usual field-strength tensor of $SU(3)_c$. Therefore, the situation is very similar to the SM, where the couplings of the $W$ boson and its Goldstone to photons and ghosts are governed by the electromagnetic gauge symmetry (i.e. the electric charge of the $W$) and a knowledge of the whole SM gauge group is not necessary. Thus, the VLQ (and the corresponding ghosts) couples in the same way to gluons as the photon to the $W$ (and its ghosts) with the replacement $e\to -g_s T_a$.

As mentioned, the aim of this work is to calculate QCD corrections to the Wilson coefficients appearing in Eqs. (\ref{LQeffL}) and (\ref{LSM}). To fix the order $\alpha_s$ pieces of these coefficients,
we calculate the scattering amplitude $A$ for the process
$q \bar \ell\to q \bar \ell$ both in the full theory and in the effective theory. Within the full theory we have to calculate the following ingredients: the LQ self-energy, the box diagrams and the genuine vertex corrections (see Fig.~\ref{Feynman}).
Within the effective theory we only have to calculate the genuine vertex correction.

Since the Wilson coefficients of our dimension six operators do not depend on the momenta and masses of the external particles, we put them to zero in our calculation. By doing so, we also avoid the generation of terms which correspond to operators of dimension higher than six. Similarly, this means that we also set the fermion masses in the couplings of Goldstone bosons to zero. Therefore, box diagrams or vertex corrections involving Goldstones vanish and merely their effect in the LQ self-energy remains. In our computational framework infrared (IR) divergences related to soft and collinear gluons are dimensionally regularized, manifesting themselves as $1/\varepsilon_{\rm IR}$ poles. For the gluon we use a general gauge with gauge parameter $\xi$ while for the vector LQ we use Feynman gauge (i.e. $\xi_{LQ}=1$).

In both the full and the effective theory we perform the necessary renormalizations leading to ultraviolet finite expressions for the amplitude $A$, from which we then can extract the QCD corrections to the Wilson coefficients.

We are aware of the fact that our calculation of the QCD corrections to the Wilson coefficients presented in the following subsections could be partially abbreviated at several places. For didactical reasons, however, we calculate the complete renormalized amplitude for both the full and the effective theory (for the very simple configuration of external states as stated above), mainly because we want to illustrate that the $\xi$-dependence drops out at the level of the renormalized amplitudes.

\subsection{Calculation in the Full Theory}

In the full theory the result for the $q \bar{\ell} \to  q \bar{\ell}$ amplitude (discarding terms of order $1/M^3$ and higher) can be written in lowest order as
\begin{equation}
\begin{aligned}
 A_{\rm tree}^{S}&= i \left(\! {\frac{{{{\left| {\Gamma _L^S} \right|}^2}}}{{{M^2}}}
    \langle\widetilde O_S^{LR} \rangle+ \frac{{\Gamma _L^S\Gamma
        _R^{S*}}}{{{M^2}}} \langle\widetilde O_S^{LL} \rangle  + L
    \leftrightarrow R{\mkern 1mu} } \!\right) \, ,
\\
 A_{\rm tree}^{V}&= i \left(\! - {\frac{{{{|\Gamma
        _L^V|^2}}}}{{{M^2}}} \langle\widetilde O_V^{LL} \rangle - \frac{{\Gamma
        _L^V\Gamma _R^{V*}}}{{{M^2}}} \langle \widetilde O_V^{LR}  \rangle + L
    \leftrightarrow R{\mkern 1mu} } \!\right) \, ,
\end{aligned}
\end{equation}
for scalar and vector LQ exchange, respectively.
The symbol $\langle \tilde O \rangle$ is a short-hand notation for the tree-level matrix element
$\langle \bar \ell q | \tilde O | q \bar \ell \rangle$ associated with the operators in (\ref{LQeffL}). In the following we calculate order $\alpha_s$ QCD corrections to this amplitude, discussing in turn the contributions due to the LQ self-energy, the vertex corrections and the box diagram.

We identify the corresponding self-energy diagram in Fig.~\ref{Feynman} (with amputated external legs) with $-i \Sigma_S(p^2)$ for scalar LQs. For working out its direct contribution to the amplitude $A$, we need $\Sigma_S$ at $p^2=0$. In our computation we also have to renormalize the mass $M$ of the leptoquark, which we do in the on-shell scheme. As the corresponding renormalization constant is directly related to  $\Sigma_S(M^2)$, we give the results at $p^2=M^2$ and at $p^2=0$, reading
\begin{eqnarray}
\begin{aligned}
\dfrac{\Sigma_S \left( {{M^2}} \right)}{{M^2}} &= \frac{\alpha
  _s}{4\pi}{C_F}\left({\frac{3}{\varepsilon }+3 l_\mu + 7}\right)\,,\\
\dfrac{\Sigma_S \left( 0 \right)}{{M^2}}&= \frac{\alpha _s}{4\pi}{C_F}\xi
\left({\frac{1}{\varepsilon }+ l_\mu + 1}\right) \,,
\end{aligned}
\end{eqnarray}
with $l_\mu=\log(\mu^2/M^2)$.

The combined effect of the direct contribution and the renomalization constant of the LQ mass leads to the occurrence of $\Sigma_S(M^2)-\Sigma_S(0)$ at the level of the amplitude $A$.  Therefore, we only kept self-energy bubble diagrams in the above expressions, i.e. all self-energy contributions  which are not tadpoles, as the latter drop out in the difference.

For the vector LQ we identify the corresponding diagram with $+i \Sigma_V^{\mu \nu}$. In our computation we only need the part proportional to $+i g_{\mu\nu}$ which we denote as $\Sigma_V(p^2)$. Again, the expressions for $p^2=0$ and $p^2=M^2$ are needed, reading
\begin{eqnarray}
\begin{aligned}
\dfrac{\Sigma_V \left( {{M^2}} \right)}{M^2}\!\! &= \frac{\alpha
  _s}{36\pi}C_F\left(\frac{57}{\varepsilon}+57 l_\mu +89 \right)\,,\\
\dfrac{\Sigma_V \left( 0 \right)}{M^2}\! &=  \frac{
  \alpha _s}{4 \pi   } C_F\left(\!\!\frac{(\xi +5)}{2}
  \left(\!\frac{1}{\varepsilon}+\!  l_\mu \right)+\frac{(\xi \!+\!7)}{4} \! \right)\,.
  \end{aligned}
\end{eqnarray}
The vertex corrections lead to the following contribution to the amplitude

\begin{eqnarray}
\begin{aligned}
A^{S,V}_{\rm tree}\left(1+ 2 \Lambda_{S,V}\right)\,,
\end{aligned}
\end{eqnarray}
with
\begin{align}
\Lambda_S  &= \frac{\alpha _s}{4\pi}{C_F}\xi \left({\frac{1}{\varepsilon }+ l_\mu + 1}\right)\,,
\end{align}
\smallskip
\begin{align}
\Lambda_{V}&=\frac{ \alpha _s }{16 \pi}C_F(\xi +3)
\left(\frac{3}{\varepsilon}+3 l_\mu +\frac{5}{2} \right)\,.
\end{align}
The box diagram contribution to the amplitudes can be compactly written as
\begin{widetext}
\begin{eqnarray}
\begin{aligned}
\Delta_S  &= \frac{\alpha_s}{4\pi} C_F  \,k_1 \, (1-\xi) \, A^S_{\rm tree} -
\frac{i}{M^2}\frac{\alpha _s}{16\pi}{C_F} \, k_2 \,
\left[  \, \left( \left|
          {\Gamma^S _L} \right|^2  \langle\widetilde
    O_{2\gamma}^{LR}\rangle + {\Gamma _L^{S*}{\Gamma^S _R}\langle{\widetilde
          O}^{RR}_{2\gamma} \rangle} \right) +L \leftrightarrow R \right] \,,\\
\Delta_V  &= \frac{\alpha_s}{4\pi} C_F  \,k_1 \, (1-\xi) \, A^V_{\rm tree} +
\frac{i}{M^2}\frac{\alpha _s}{16\pi}{C_F} \, k_2 \,
\left[ \, \left( \left|
          {\Gamma^V _L} \right|^2  \langle\widetilde
    O_{3\gamma}^{LL}\rangle + {\Gamma _L^{V}{\Gamma^{V*} _R}\langle{\widetilde
          O}^{LR}_{3\gamma} \rangle} \right) +L \leftrightarrow R \right] \,,
\label{ScalarBox}
\end{aligned}
\label{box}
\end{eqnarray}
\end{widetext}
for scalar and vector LQs, respectively. The expressions for $k_1$ and $k_2$ read
\begin{equation}
\begin{aligned}
k_1 &= \frac{1}{\varepsilon_{\rm IR}} + 1 + l_\mu\,,\qquad
k_2 &=  \frac{1}{\varepsilon_{\rm IR}} + \frac{3}{2} + l_\mu \, .
\end{aligned}
\end{equation}
The symbols $\langle \widetilde O \rangle$ again denote the tree-level matrix
elements of the operators $\widetilde O$ in (\ref{LQeffL}). Furthermore, we defined
\begin{eqnarray}
\begin{aligned}
&&\langle \widetilde O^{AB}_{2\gamma } \rangle = \langle \bar q{\gamma ^\mu
}{\gamma ^\nu }{P_A}\bar \ell \ell {\gamma
  _\nu }{\gamma _\mu }{P_B}q \rangle \, ,  \\
&&\langle \widetilde O^{AB}_{3\gamma } \rangle = \langle \bar q{\gamma ^\mu
}{\gamma ^\nu }{\gamma ^\sigma }{P_A}\bar \ell \ell {\gamma _\sigma }{\gamma
  _\nu }{\gamma _\mu }{P_B}q \rangle \, .
\end{aligned}
\end{eqnarray}

In addition to the contributions of the various diagrams just described, we have to renormalize the coupling constants of the leptoquark to the fermions and we have to take into account the LSZ factor of the quark fields. Thereby the couplings get replaced by  $\Gamma _{L,R}^{S,V} \to \Gamma _{L,R}^{S,V}\left( {1 + \delta
\Gamma _{L,R}^{S,V}} \right)$ and the quark fields by $q \to q \left( {1 +
\delta Z_q} \right)$. The explicit expressions read
\begin{equation}
\begin{aligned}
\delta \Gamma _{L,R}^S &=  - \frac{\alpha _s}{4\pi} C_F
\frac{3}{2} \frac{1}{\varepsilon}
\,,\qquad
\delta \Gamma _{L,R}^V &=  - \frac{\alpha _s}{4\pi} C_F \frac{25}{6} \frac{1}{\varepsilon}
   \, ,  \\
\delta Z_q &=  \frac{\alpha_s}{4\pi} C_F \, \xi \, \left(
  \frac{1}{\varepsilon_{\rm IR}} - \frac{1}{\varepsilon} \right) \, .
  \end{aligned}
\end{equation}
Note
that we have already discussed (and taken into account) the renormalization of the mass of the LQ.
The final renormalized results for the amplitudes in the full theory read
\begin{widetext}
\begin{eqnarray}
A_{\rm full,ren}^{S(V),\alpha_s}&=\Delta_{S(V)} \! + \! \left( {2\Lambda_{S(V)}  + \dfrac{{\Sigma_{S,(V)}
        \left( {{M^2}} \right)  \! -  \! \Sigma_{S(V)} \left( 0 \right)}}{{{M^2}}}}
  + \delta Z_q + 2 \delta \Gamma _{L,R}^{S(V)}
  \!\right) \,   A^{S(V)}_{\rm tree} \, .
\label{resultfull}
\end{eqnarray}
\end{widetext}

When inserting all the ingredients listed above into these formulas, we see that the ultraviolet singularities are cancelled. Also the $\xi$ dependence
related to the gluon cancels in these expressions, as it should be the case when calculating on-shell matrix elements. Only $1/\varepsilon_{\rm IR}$
infrared singularities remain which will enter the corresponding matrix element in the effective theory in precisely the same way, leading to finite Wilson coefficients.

\subsection{Effective Theory and Matching}

Within the effective theory, we first calculate the QCD corrections to the amplitudes $q \bar \ell\to q \bar \ell$ originating from the operators in~\eq{LQeffL} (see last diagram in FIG.~\ref{Feynman}). We obtain
\begin{widetext}
\begin{eqnarray}
\begin{aligned}
\Delta_S^{\rm eff}  &= \frac{\alpha_s}{4\pi} C_F  \left(
 \frac{1}{\varepsilon_{\rm IR}} -\frac{1}{\varepsilon} \right) \, (1-\xi) \, A^S_{\rm tree} -
\frac{i}{M^2}\frac{\alpha _s}{16\pi}{C_F} \, \left(
  \frac{1}{\varepsilon_{\rm IR}} -\frac{1}{\varepsilon} \right)  \,
\left[  \, \left( \left|
          {\Gamma^S _L} \right|^2  \langle\widetilde
    O_{2\gamma}^{LR}\rangle + {\Gamma _L^{S*}{\Gamma^S _R}\langle{\widetilde
          O}^{RR}_{2\gamma} \rangle} \right) +L \leftrightarrow R \right] \,,\\
\Delta_V^{\rm eff}  &= \frac{\alpha_s}{4\pi} C_F  \,  \left(
  \frac{1}{\varepsilon_{\rm IR}} -\frac{1}{\varepsilon} \right)  \, (1-\xi) \, A^V_{\rm tree} +
\frac{i}{M^2}\frac{\alpha _s}{16\pi}{C_F} \,  \left(
  \frac{1}{\varepsilon_{\rm IR}} -\frac{1}{\varepsilon} \right) \,
\left[ \, \left( \left|
          {\Gamma^V _L} \right|^2  \langle\widetilde
    O_{3\gamma}^{LL}\rangle + {\Gamma _L^{V}{\Gamma^{V*} _R}\langle{\widetilde
          O}^{LR}_{3\gamma} \rangle} \right) +L \leftrightarrow R \right] \,,
\label{EFTBox}
\end{aligned}
\end{eqnarray}
\end{widetext}
for scalar and vector LQs, respectively. We immediately see that the $\xi$-dependence drops out when taking into account the effect of $\delta Z_q$. Furthermore, we observe that the infrared divergences are then the same as in the full theory. Since we are interested in the matching, we drop at this level all the terms involving  $1/\varepsilon_{\rm IR}$ in both versions of the theory. We then rewrite
\begin{equation}
\widetilde
O_{2\gamma}^{AB}=(4-2\varepsilon) \widetilde O_S^{AB}+ \widetilde
O_T^{AB} \,.
\end{equation}
$\widetilde O_T^{AB}$ vanishes in $d=4$ dimensions for $A \neq B$. Therefore, it plays the role of an evanescent operator. For $A = B$ we have $\widetilde O_T^{AA}=\widetilde O_T^{A}$, where $\widetilde O_T^{A}$ is present in the operator basis.
\newline
Furthermore, we rewrite $\widetilde O_{3\gamma}^{AB}$ (see
e.g. Ref.~\cite{Chetyrkin:1997gb}) as
\begin{equation}
\begin{aligned}
\widetilde O_{3\gamma }^{LL} &=  \left( {4 -
    8\varepsilon } \right)\widetilde O_V^{LL} - \widetilde O_E^{LL} \,,\\
\widetilde O_{3\gamma }^{LR} &=  16\left( {1 -
    \varepsilon } \right)\widetilde O_V^{LR} - \widetilde O_E^{LR} \,,
\end{aligned}
\label{O3g}
\end{equation}
where  $\widetilde O_E^{LL}$ and  $\widetilde O_E^{LR}$ are evanescent operators. Renormalizing the operators (including the evanescent ones) in the $\overline{\rm MS}$-scheme, we obtain (after dropping the infrared terms as described)
\begin{widetext}
\begin{align}
\begin{aligned}
A_{\rm eff,ren}^{S,\alpha_s} = &-\frac{i}{M^2} \frac{\alpha_s}{8\pi} C_F
\left[ |\Gamma_L^S|^2 \langle \widetilde O_S^{LR} \rangle + \Gamma_L^S
  \Gamma_R^{S*} \langle \widetilde O_S^{LL} \rangle \right]  + i \delta \widetilde C_S^{LR} \langle \widetilde O_S^{LR} \rangle +
i \delta \widetilde C_S^{LL} \langle \widetilde O_S^{LL} \rangle +
i \delta \widetilde C_T^{R} \langle \widetilde O_T^{R} \rangle + (L
\leftrightarrow R) \, ,\\
A_{\rm eff,ren}^{V,\alpha_s} &= \frac{i}{M^2} \frac{\alpha_s}{2\pi} C_F
\left[ |\Gamma_L^V|^2 \langle \widetilde O_V^{LL} \rangle + 2 \Gamma_L^V
\Gamma_R^{V*} \langle \widetilde O_V^{LR} \rangle \right] +i \delta \widetilde C_V^{LL} \langle \widetilde O_V^{LL} \rangle
+ i \delta \widetilde C_V^{LR} \langle \widetilde O_V^{LR} \rangle +
(L \leftrightarrow R) \, .
\end{aligned}
\end{align}
\end{widetext}
The quantities $\delta C$ contain the order $\alpha_s$ corrections to the respective Wilson coefficients. The corresponding results in the full theory (also after dropping the infrared terms) read

\begin{widetext}
\begin{align}
A_{\rm full,ren}^{S,\alpha_s} &= \frac{i}{M^2} \frac{\alpha_s}{4\pi} C_F \left(3 l_\mu +\frac{13}{2}\right)
\left[ |\Gamma_L^S|^2 \langle \widetilde O_S^{LR} \rangle +\, \Gamma_L^S \Gamma_R^{S*} \langle \widetilde O_S^{LL} \rangle \right] -\frac{i}{M^2} \frac{\alpha_s}{16\pi} C_F \left(l_\mu +\frac{3}{2}\right) \Gamma_L^{S*}
\Gamma_R^S \langle \widetilde O_T^{R} \rangle
 + (L \leftrightarrow R) \, , \nonumber\\
 A_{\rm full,ren}^{V,\alpha_s} &= - \frac{i}{M^2} \frac{\alpha_s}{72\pi} C_F \left[5 \left(30 l_\mu +41
\right)
 |\Gamma_L^V|^2 \langle \widetilde O_V^{LL} \rangle   +4 \left(24 l_\mu +31\right) \Gamma_L^V
 \Gamma_R^{V*} \langle \widetilde O_V^{LR} \rangle \right]
 + (L \leftrightarrow R) \, .
\end{align}
\end{widetext}
From these equations the Wilson coefficients are easily determined, reading
\begin{eqnarray}
\begin{aligned}
\widetilde C^{LR}_S &= \frac{{\left| {\Gamma _L^S} \right|^2}}{{{M^2}}}\left( {1 + \frac{{{\alpha _s}}}{{4\pi }}C_F} \left( {{\rm{3}}l_\mu{\rm{ + 7 }}} \right)\right),\\
\widetilde C^{RR}_S &= \frac{{\Gamma _R^S\Gamma _L^{S*}}}{{{M^2}}}\left( {1 + \frac{{{\alpha _s}}}{{4\pi }}C_F} \left( {{\rm{3}}l_\mu{\rm{ + 7  }}} \right)\right),\\
{{\widetilde C}}^R_T &=  - \frac{{\Gamma _R^{S}\Gamma _L^{S*}}}{{{M^2}}}\frac{{{\alpha _s}}}{{16\pi }}C_F\left( {l_\mu{{ + \frac{3}{2}}}} \right)\,,\\
{\widetilde C}^{LL}_V &=  - \frac{{\left| {\Gamma _L^V} \right|^2}}{{{M^2}}}\left( {1 + \frac{{{\alpha _{s}}}}{{72\pi }}{{{C}}_{{F}}}\left( {150l_\mu + 241} \right)} \right)\,,\\
{\widetilde C}^{LR}_V &=  - \frac{{\Gamma _L^{V}\Gamma _R^{V*}}}{{{M^2}}}\left( {1 + \frac{{{\alpha _{{s}}}}}{{18\pi }}{{{C}}_{{F}}}\left( {24l_\mu + 49} \right)} \right)\,,
\end{aligned}
\end{eqnarray}
and the corresponding equations obtained by exchanging $L$ and $R$.

\begin{figure*}[t]
	\begin{center}
		\begin{tabular}{cp{7mm}c}
			\includegraphics[width=0.56\textwidth]{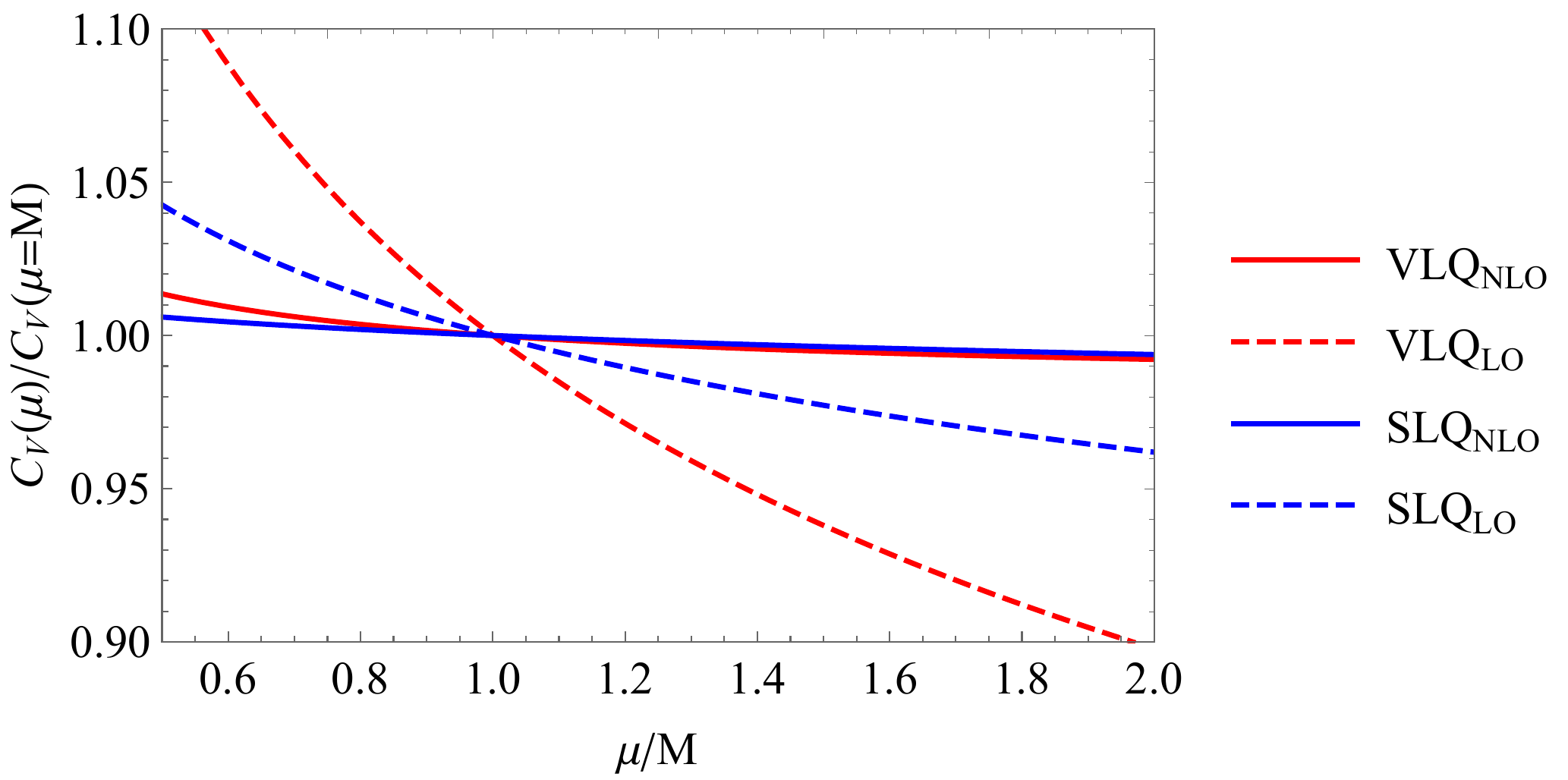}
			\includegraphics[width=0.43\textwidth]{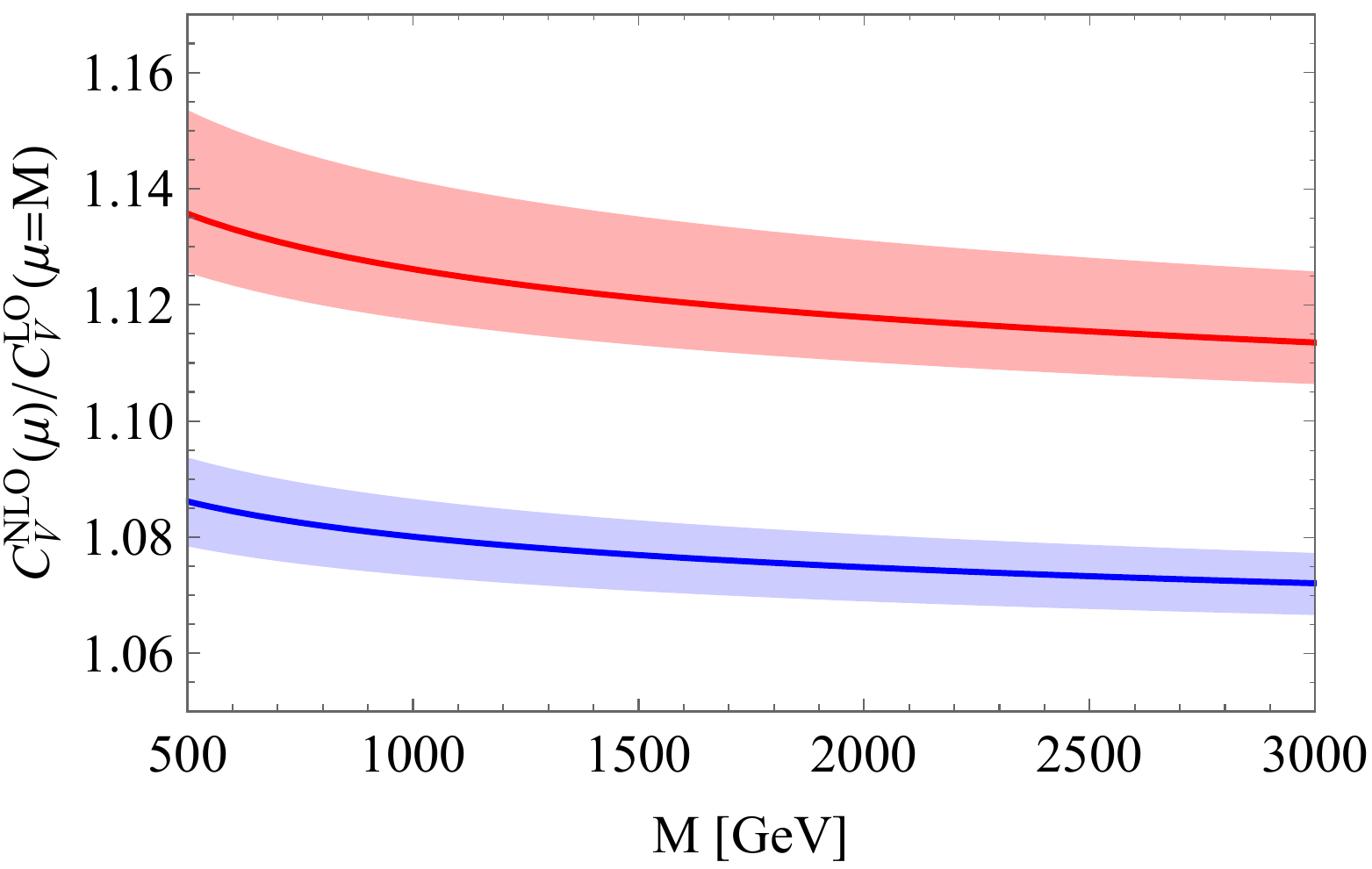}
		\end{tabular}
	\end{center}
	\caption{Left: Dependence of the Wilson coefficients of vector operators (in the SM basis) on a variation of the matching scale $\mu$ for $M=1\,{\rm TeV}$. Note that the ratio $C_V(\mu)/C_V(\mu=M)$ only depends very weakly on the overall scale $M$. One can clearly see that the scale dependence significantly reduces when
		going from LO to NLO. The $\mu$ dependence of the vector operator is bigger when generated by a VLQ rather than by a SLQ.
	Right: Ratio of the Wilson coefficient of vector operators  calculated at NLO in $\alpha_s$ ($C_{V}^{\rm NLO}$) to the corresponding Wilson coefficient at leading order at $\mu=M$ ($C_{V}^{\rm LO}(\mu=M)$). The corresponding coloured regions indicate the remaining NLO matching scale uncertainties and are obtained by varying $\mu$ between $1/2M$ and $2M$.}
	\label{FigNum}
\end{figure*}

Finally, we have to discuss the transition to the SM basis (\eq{LSM}), in
which the QCD running and the physical observables are calculated. A naive
four-dimensional Fierz
transformation is not applicable \`a priori at the one-loop level.
Instead, we require that the renormalized matrix elements for the process $q \bar
\ell \to q \bar \ell$ calculated in both versions of the effective theory
coincide. Due to the specific definition of the evanescent operators
(see (Eq.~\ref{O3g})), we obtain for the Wilson coefficients
\begin{eqnarray}
\begin{aligned}
C_V^{LL,RR} &= \widetilde C_V^{LL,RR}\,,\\
C_S^{LR,RL} &=  - 2\widetilde C_V^{RL,LR}\,,
\end{aligned}
\end{eqnarray}
i.e., there are no corrections with respect to the 4-dimensional Fierz
identities for vector LQs. However, for the operators generated by scalar LQs
this is not the case and corrections to the naive Fierz identities do appear,
as our results show:
\begin{eqnarray}
\begin{aligned}
C_V^{RL} &=  - \frac{1}{2}\widetilde C_S^{LR}\left( {1 + \frac{3}{{8\pi }}{C_F}{\alpha _s}} \right)\,,\\
C_S^{LL} &=  - \frac{1}{2}\widetilde C_S^{LL}\left( {1 + \frac{7}{{8\pi }}{C_F}{\alpha _s}} \right) - 6\widetilde C_T^L\,,\\
C_T^{LL} &=  - \frac{1}{8}\widetilde C_S^{LL}\left( {1 - \frac{1}{{8\pi }}{C_F}{\alpha _s}} \right) + \frac{1}{2}\widetilde C_T^L\,.
\end{aligned}
\end{eqnarray}

At this point, we should remind the reader that the corrections for the vector LQs are independent of the gluon gauge parameter $\xi$ but depend on the LQ gauge parameter (which we set to 1 in Feynman gauge). This is an artefact of our simplified model framework for the vector LQ. In a UV complete model there will of course be additional contributions which are not taken into account in our analysis. However, also for collider analyses simplified models are used. Furthermore, since we calculate QCD corrections (and not LQ corrections) to the matching, the independence on the gluon gauge parameter is sufficient  to justify that our results are reasonable. I.e. we calculated the minimal QCD matching effects which are present in any model and only get supplemented by additional effects depending on the UV completion.

\section{Impact and Conclusions}\label{conclusions}

The results of our calculation above can be summarized in the following compact way: In the SM basis, the lowest order Wilson coefficients of vector and scalar operators receive a shift
\begin{align}
\begin{aligned}
C_V^{\rm VLQ} \to C_V^{\rm VLQ}\left( {1 + \frac{{{\alpha _{s}}}}{{72\pi }}{{{C}}_{{F}}}\left( {150l_\mu + 241} \right)} \right)\,,\\
C_S^{\rm VLQ} \to C_S^{\rm VLQ}\left( {1 + \frac{{{\alpha _{{s}}}}}{{18\pi }}{{{C}}_{{F}}}\left( {24l_\mu + 49} \right)} \right)\,,
\end{aligned}
\end{align}
if they originate from vector LQs. Concerning operators arising from scalar LQs we have
\begin{align}
C_V^{\rm SLQ} &\to C_V^{\rm SLQ}\left( {1 + \frac{{{\alpha _s}}}{{4\pi }}{C_F}\left( {{\rm{3}}l_\mu{\rm{ + }}\frac{{17}}{2}} \right)} \right)\,,\nonumber\\
C_S^{\rm SLQ} &\to C_S^{\rm SLQ}\left( {1 + \frac{{3{\alpha _s}}}{{2\pi }}{C_F}} \right)\,,\\
C_T^{\rm SLQ} &\to C_T^{\rm SLQ}\left( {1 + \frac{{{\alpha _s}}}{\pi }{C_F}\left( {l_\mu{\rm{ + 2}}} \right)} \right)\,,\nonumber
\end{align}
for all Wilson coefficients of vector, scalar and tensor operators, respectively. Note that these formulas are valid for all 10 representations of scalar and vector LQs which have couplings invariant under the SM gauge group~\cite{Buchmuller:1986zs} to quarks and leptons and even apply if the LQ couples to right-handed neutrinos instead of SM leptons. They also do not depend on whether or not charged conjugated fields are involved, since charge conjugation can only lead to a change in the chirality and/or to an overall relative minus sign and therefore does not affect the results given above.

Let us now consider the numerical impact of the corrections we calculated. Here we focus on the phenomenologically most important case of vector operators (in the SM basis) since they are capable of explaining both the tensions in $b\to c\tau\nu$ processes and in $b\to s\mu^+\mu^-$ observables (see e.g. Ref.~\cite{Crivellin:2018gzw} for a recent overview). Note that since vector operators in the SM basis are conserved under QCD, the dependence of their Wilson coefficients on the scale $\mu$ can be directly identified with the matching scale uncertainty once the running of the couplings $\Gamma^{S,V}$ is taken into account.

First, we examine the dependence on the matching scale $\mu$ in the left-handed plot of Fig.~\ref{FigNum}. For this purpose we show the ratio
$C_V(\mu)/C_V(\mu=M)$ both at LO and at NLO. For the LO estimate we only kept the implicit scale dependence via the couplings
\begin{equation}
\Gamma \left( \mu  \right) = \Gamma \left( {{\mu _0}} \right){\left( {\frac{{{\alpha _s}\left( \mu  \right)}}{{{\alpha _s}\left( {{\mu _0}} \right)}}} \right)^{\frac{{\gamma _\Gamma ^0}}{{2{\beta _0}}}}}\,,
\end{equation}
with $\beta_0=7$ and $\gamma _{{\Gamma _S}}^0 = 3 C_F$, $
\gamma _{{\Gamma _V}}^0 = \frac{{25}}{3}C_F$ for scalar and vector LQs, respectively\footnote{Note that for $\beta_0$ we used the SM value with 6 active flavours. In principle, this changes in the full theory depending on the number of LQ representations added. However, the specific value of $\beta_0$ does not affect the cancellation of the leading matching scale uncertainty. }. It can be clearly seen that the scale uncertainty is significantly reduced by the NLO corrections compared to the LO estimate.

Now let us consider the numerical impact of our NLO calculation. Here we show the ratio $C_{V}^{\rm NLO}(\mu)/C_{V}^{\rm LO}(\mu=M)$ in the right plot of Fig.~\ref{FigNum}. I.e. we show the relative effect of the NLO correction with respect to the naive tree-level result. The NLO correction is constructive, meaning that the size of the Wilson coefficients is increased by around 8 \% (13 \%) if they originate from scalar (vector) LQs. The remaining matching scale uncertainty at NLO is indicated by the coloured bands obtained by varying $\mu$ (encoded in $C_{V}^{\rm NLO}(\mu)$) from $M/2$ to $2M$.

Thus, assuming that LQs account for the discrepancies between the measurement and the SM predictions in semi-leptonic $B$ decays, the mass can be larger (assuming a fixed coupling) than without including the QCD corrections to the matching. This means that signal strength in LHC searches is reduced, increasing the allowed parameter space of the models.

Finally, note that even though these operators are very important for explaining the hints for NP in these observables, our results are not at all limited to this class of processes but evidently also apply to e.g. (semi) leptonic Kaon decays, tau decays and even neutralino DM matter scattering in the MSSM where the squark (neutralino) takes the role of the scalar LQ (lepton).

\acknowledgements

The work of C. G. is supported by the Swiss National Foundation under grant 200020\_175449/1. The work of A. C. is supported by an Ambizione Grant of the Swiss National Science Foundation (PZ00P2\_154834). The  work  of J. A.  is  supported  by  the  DFG  cluster  of  excellence ``Origin and Structure of the Universe''. A. C. is grateful to Michael Spira and Yannick Ulrich for useful discussion. J. A. thanks Christoph Bobeth for useful discussions and C.G. thanks Massimo Passera for useful discussions.

\bibliography{BIB}

\end{document}